\documentclass[12pt]{article}
\usepackage[utf8]{inputenc}
\usepackage[T1]{fontenc}
\usepackage{graphicx}
\usepackage{epstopdf} 
\usepackage{color}
\input{epsf}
\newcommand{\be}{\begin{equation}}
\newcommand{\ee}{\end{equation}}
\newcommand{\bear}{\begin{eqnarray}}
\newcommand{\ear}{\end{eqnarray}}
\renewcommand{\theequation}{\arabic{section}.\arabic{equation}}
\newcommand{\wt}{\widetilde}

\begin{document}
\title{Kelvin transformation and inverse multipoles in electrostatics}
\author{R. L. P. G. Amaral,$^*$  O. S. Ventura$^{\dagger}$ and N. A. Lemos$^*$\\
\small
$^*${\it Instituto de F\'{\i}sica - Universidade Federal Fluminense}\\
\small
{\it Av. Litor\^anea, S/N, Boa Viagem, Niter\'oi,
CEP.24210-340, Rio de Janeiro - Brasil}\\
\small
$^{\dagger}${\it Centro Federal de Educa\c{c}\~ao Tecnol\'ogica do Rio de
Janeiro}\\
\small
{\it Av.Maracan\~a 249, 20271-110, Rio de Janeiro - RJ, Brazil}\\
\small
{\it rubens@if.uff.br, ozemar.ventura@cefet-rj.br and nivaldo@if.uff.br}}

\date{\today}

\maketitle
\begin{abstract}
	
The  inversion in the sphere or Kelvin transformation, which exchanges the radial coordinate for its inverse, is used as a guide
to relate distinct electrostatic problems with dual features. The exact solution of some nontrivial problems are obtained through the mapping from simple highly symmetric systems. In particular, the concept of multipole expansion  is revisited from a point of view opposed to the usual one: the sources are distributed in
a region far from the origin while the electrostatic potential is described at points close to it. 

\end{abstract}

%%%%%%%%%%%%%%%%%%%%%%%%%%%%%%%%%%%%%%%%%%%%%%%%%%%%%%%%%%%%%%%%%%%%
\section{Introduction}

Mapping a difficult problem into an easier or previously solved one is a powerful strategy both in mathematics and physics. Electromagnetism is a theory in which this is often possible and rewarding. For example, two arbitrary charge distributions  together with their respective electrostatic potentials are related by Green's reciprocation theorem, allowing known results about a simple arrangement of charges to be translated into  information about a more complex configuration \cite{jackson}. Another useful technique in electrostatics is the Kelvin transformation \cite{Kelvin} --- also known as inversion in the sphere  ---   which, among other things, maps planes into spheres and vice versa, and by means of which some difficult problems can be solved \cite{Jeans,Smythe}. Such mappings are often suggested by symmetries of the theory. Here we explore certain aspects of the application of the Kelvin transformation to electrostatics that we find instructive and, to our knowledge, have not been discussed 
elsewhere.

Electromagnetism is an example of a successful theory with impressive experimental corroboration both at classical and quantum realms. Lorentz  invariance insures its validity at high velocities, while the gauge symmetry  establishes a paradigm for the description of other fundamental interactions and is linked to charge conservation and the absence of photon mass. The latter aspect guarantees the scale invariance of the electromagnetic theory. A more subtle property of the theory, intimately related to scale invariance, is its conformal invariance.
As has been recently stressed \cite{jackiw1}, the conformal symmetry of electromagnetism is characteristic to  four dimensional space-time.

Within the conformal symmetry group, the special conformal transformation
means a spacetime coordinate inversion followed by a spacetime translation and another inversion,
\begin{equation}
\frac{x_\mu}{x^2}= { \frac{\wt x_\mu}{{\wt x}^2}}
+b_\mu,
\label{aq}
\end{equation}  
and takes the form
\begin{eqnarray}
\label{special-conformal-transformation-spacetime}
	x_\mu= {\frac {{\wt x_\mu} +{\wt x^2}b_\mu}{1+2b \cdot {\wt x} +b^2{\wt x}^2}}\, ,
	\label{aq1}
\end{eqnarray} 
where $a \cdot b = a_{\mu}b^{\mu}$ and $a^2 = a_{\mu}a^{\mu}$.
Its implications 
to the dynamics of charges have been discussed in \cite{invconfor} and 
a more complete investigation of this symmetry is presented in \cite{cohn}, where   use is made of the 
general covariant formalism.

This article deals with  conformal transformations  analogous to Eq. (\ref{aq1}) which affect spatial variables alone, so that the time variable is left untouched. In particular it will be focused on an essential ingredient of the  special conformal transformation, namely
the spatial inversion
%, which will be taken only in the spatial coordinates:
\be\label{si-index}
{\widetilde x_i}=\frac{R^2}{r^2} x_i \, ,
\label{4in}
\ee
where $R$ is a positive constant an $r=\sqrt{{\bf x}\cdot \textbf{x}}$ is the radial variable.
This mapping, variously known as inversion in the sphere or Kelvin transformation \cite{Kelvin},  leaves electrostatics invariant, the focus of our interest. It is worth remarking that, in general,  magnetostatics is not left invariant 
by (\ref{si-index}).
This kind of coordinate change has been explored 
in the framework of electrical engineering \cite{eng}, but some of its
features have  not been  appreciated  from the physicist's point of view. 
For instance, it allows the description of infinitely extended systems
starting from localized ones. 

An important tool for the study of localized charge distributions  is
the multipole expansion, which has been widely explored not only in electromagnetism but also in other
 macroscopic field theories such as gravitation. In the latter case, the study of Newtonian and Einsteinian orbits 
is an explicit example  \cite{multipolograv,gravitation}.
Perturbations of the Newtonian gravitational potential imply  planetary perihelion advance. 
In this case, the decisive  perturbations, which stem from space-time curvature, 
where once thought to be due  to a solar oblateness
that would give rise to a quadrupole contribution to the Sun's gravitational potential  \cite{orbitasmultipolos,etmultipolos}.
In the case of electrostatics, the standard textbooks devote great attention to the multipole expansion \cite{jackson2}. Its applications are wide ranging, from the quantum-mechanical study of asymmetric atoms \cite{multipoloatomo} up to electromagnetic radiation and scattering of electromagnetic waves
\cite{jackson3}. As a rule, one is interested in  describing a field at points far from  a localized source distribution,  as, for example,  in the discussion of the electric field created by a point dipole or by a uniformly polarized spheroidal electret embedded in an infinite dielectric \cite{amaralnivaldo}. An unusual point that will be addressed here is the transformation of the multipole expansion
of the electrostatic potential under the inversion in the sphere (\ref{si-index}). This gives rise to  an interchange of the roles of points close to
and distant from the origin.  

The paper is  organized as follows. Section 2 deals with the effect on the Poisson equation of inversions in the sphere. Their impact on the multipole expansion of the electrostatic potential is investigated in  Section 3. The following sections are dedicated to  applications. In Section 4
the duality of spherical shells leads naturally to the concept of self-dual and anti-self-dual models, and to the role they play in the method of images.  In Section 5 the consideration of eccentric spheres leads to the discussion of a general conformal transformation. In Section 6 the mapping from spheres into planes is discussed stressing the topology change induced by the inversion transformation. In section 7 the relationship between a cylinder and a special torus is studied. Finally,  some conclusions are presented and further applications are pointed out.

% % % % % % % % % % % % % % % % % % % % %

\section{Electrostatics and Inversion Transformation}
\setcounter{equation}{0}

Let us start by considering the role of the inversion in the sphere 
\be\label{si}
{\bf r} \,\, \stackrel {{\cal S}}{ \longrightarrow } \,\, {\widetilde {\bf r}}=  \frac{R^2}{r^2} \, {\bf r} 
\ee
in electrostatics.\footnote{The positive parameter $R$ is required for dimensional consistency and  defines the radius of an invariant sphere. A change in $R$ means a scale transformation.} Since all information on the electrostatic field is embodied in the  potential  $\Phi$, all is needed is a description of its fate under transformation (\ref{si}), which is denoted by $\cal S$ and whose inverse is
\be\label{si-inverse}
%{\wt {\bf r}} \,\, \stackrel {{\cal S}^{-1}}{ \longrightarrow } \,\,  
{\bf r}=  \frac{R^2}{{\wt r}^2}\, {\widetilde{\bf r}} \, .
\ee

In order to determine how solutions of the Poisson equation
 \begin{eqnarray}\label{pe}
 \nabla^2\Phi=-\frac{1}{\epsilon_0}\rho
 \label{2}
 \end{eqnarray}  
are mapped into other solutions by the inversion operation,  we start from the Laplacian  in spherical coordinates:
\begin{eqnarray}
\label{laplancian-spherical}
\nabla^2=\frac{1}{r^2}\frac{\partial}{\partial r}\Bigl( r^2\frac{\partial}{\partial  r}\Bigr) + 
\frac{1}{r^2}\bigg[ \frac{1}{\sin \theta}\frac{\partial}{\partial \theta}\Bigl( \sin \theta \frac{\partial}{\partial \theta}\Bigr) + \frac{1}{\sin^2\theta}\frac{\partial^2}{\partial \phi^2} \bigg] = 
\frac1{r^2}\left[D_r + D_r^2+L^2\right].
\label{1}
\end{eqnarray} 
Here $D_r=r\partial /\partial r$ and  $L^2$ is a differential operator acting on the angular 
variables alone   \cite{Landim}.
The angular operator is invariant because  transformation (\ref{si})  does not change angles: 
\begin{equation}
\label{invariance-angles}
\frac{{\widetilde {\bf r}} \cdot {\widetilde {\bf r}^\prime}}{\widetilde r \, \wt r^\prime}=\frac{{\bf r} \cdot {{\bf r}^\prime}}{ r\, r^{\prime}}\, .
\end{equation}
Furthermore, taking the modulus of both sides of (\ref{si-inverse}) we find
\begin{equation}
\label{r-til-r}
{ r} = \frac{R^2}{\widetilde r} \, ,
\end{equation}
from which it follows that
\begin{equation}
\label{invariance-rDr}
D_{{\widetilde r}} = {\widetilde r}\frac{\partial }{\partial \wt r} = {\widetilde r} \frac{\partial r}{\partial {\widetilde r} } \frac{\partial}{\partial r}
 ={\wt r}\Bigl( -\frac{R^2}{{\wt r}^2}\Bigr) \frac{\partial}{\partial r} =
-  r \frac{\partial}{\partial  r} =
- D_{ r} \, .
\end{equation}
Certainly,  the term linear in $D_r$  spoils the invariance of the Laplacian operator (\ref{laplancian-spherical}) under inversions.

It is straightforward, although a little tedious, to show that we can get rid of the term linear in $D_r$ by  the following device:
\begin{eqnarray}
\nabla^2 \Phi =\nabla^2 \bigl( r^{-{1\over 2}}r^{1\over 2}\Phi\bigr)=r^{-{5\over 2}}\left[-{1\over 4}+ D_r^2+L^2\right]\bigl(r^{1\over 2}\Phi\bigr).
\label{4}
\end{eqnarray} 
The Poisson equation is thus written as
\begin{equation}
\label{Poisson-without-linear-Dr}
\left[-{1\over 4}+ D_r^2+L^2\right]r^{1\over 2}\Phi ({\bf r})=-\frac 1{\epsilon_0}r^{{5\over 2}}\rho({\bf r}) \, .
\end{equation}
 This suggests to define 
\begin{eqnarray}\label{fieldexchange}
{\wt r}^{1/2}\wt\Phi({\wt {\bf r}})= r^{1/2} \Phi ( {\bf r}) \,\, \Longrightarrow 
\,\, \wt\Phi({\wt {\bf r}}) = \left(\frac{ r}{\wt r}\right)^{1\over 2}\Phi ( {\bf r}) = \frac{ R}{{\wt r}}\Phi({ \bf r}) 
\end{eqnarray} 
and, similarly, 
\begin{eqnarray}\label{densityexchange}
{\wt r}^{5/2}{\wt \rho} ({\wt {\bf r}})= r^{5/2} \rho ( {\bf r}) \,\, \Longrightarrow 
\,\, \wt\rho ( {\wt {\bf r}}) = \left(\frac{ r}{\wt r}\right)^{5\over 2}\rho 
( {\bf r}) = \bigg( \frac{ R}{{\wt r}}\bigg)^5\rho ( {\bf r}) 
\end{eqnarray} 
with $\bf r$ given in terms of $\wt{\bf r}$ by (\ref{si-inverse}).
With the basic definitions (\ref{fieldexchange}) and (\ref{densityexchange}) the Poisson equation is preserved, giving rise to a pair of dual electrostatics problems related by the space inversion ${\cal S}$:
\begin{eqnarray}
\nabla^2\Phi({\bf r})=-\frac 1{\epsilon_0}\rho({\bf r})\,\, \stackrel{{\cal S}}{\Longleftrightarrow} \,\, {\wt\nabla}^2\wt \Phi({\wt {\bf r}})=-\frac 1{\epsilon_0}\wt\rho( {\wt {\bf r}}) \, .
\label{7}
\end{eqnarray}
Thus, we have two electrostatic problems that are derived from each another by means of the inversion transformation. The transformation (\ref{fieldexchange}) of one harmonic function into another appears in \cite{Kellog} under the name of a Kelvin transformation. 
It is worthy of note, and it is easy to check, that inversion in the sphere is an involution, that is, ${\wt{\wt{\bf r}}} = {\bf r}$, ${\wt{\wt \rho}} = \rho$ and 
${\wt{\wt \Phi}} = \Phi$.

Some subtleties deserve to be stressed. Notice that a gauge transformation $\Phi^\prime({\bf r})=\Phi({\bf r})+\Phi_0$, which does not change the electric field ${\bf E}({\bf r})$ in the original setup, affects the physics described by the dual system through the ``addition'' of a point particle of charge $\wt Q_0=4\pi R\epsilon_0\Phi_0$ at the origin, since $\wt\Phi^\prime({\wt {\bf r}})=\wt\Phi( {\wt {\bf r}})+\displaystyle \frac{\Phi_0 R}{\wt r}$. Under this perspective, the spacial inversion maps a physical system into a class of systems related by the addition of a monopole at the origin.

Before turning to our main issues of interest, let us digress a little on the mathematical origin of transformations (\ref{fieldexchange}) and (\ref{densityexchange}). With the help of (\ref{r-til-r}) it is easy to show that under transformation 
(\ref{si}) the volume element changes as follows:
\begin{equation}
\label{jacobian}
{\wt {dV}} = {\wt r}^2 d{\wt r} d\Omega   = \bigg( \frac{R}{r}\bigg)^6 r^2 dr d\Omega =  \bigg( \frac{R}{r}\bigg)^6 dV \, .
\end{equation}
We also have
\begin{equation}
\label{transformation-r-r-prime}
\vert {\wt {\bf r}} - {\wt {\bf r}}^{\prime} \vert^2 = \bigg\vert \frac{R^2}{r^2} 
{\bf r} - \frac{R^2}{{r^{\prime}}^2} 
{\bf r}^{\prime} \bigg\vert^2 = R^4\frac{\vert {\bf r} - {\bf r}^{\prime}\vert^2}{r^2 {r^{\prime}}^2} \,\, \Longrightarrow \,\, \vert {\wt {\bf r}} - {\wt {\bf r}}^{\prime} \vert =  \frac{R^2}{r r^{\prime}}\vert {\bf r} - {\bf r}^{\prime}\vert \, .
\end{equation}
These results, combined with  (\ref{fieldexchange}) and (\ref{densityexchange}),  prompt a rederivation  of  Kelvin's inversion theorem by means of a change of integration variables in the Coulomb law equation:
\begin{eqnarray}\label{coulomb}
\Phi({\bf r})=\frac 1{4\pi\epsilon_0}\int dV^{\prime}\frac {\rho({ {\bf r}^\prime})}{\vert {\bf r} -{\bf r}^{\prime}\vert}+\Phi_0 \,\,\,\, 
%\stackrel{{\cal S}}
{\Longleftrightarrow} \,\,\,\, \wt\Phi({\wt {\bf r}})=
\frac 1{4\pi\epsilon_0}\int \wt{d {V^\prime}}\frac {{\wt\rho}({ {\wt{\bf r}}^\prime})}{\vert {\wt {\bf r}} -{\wt {\bf r}}^{\prime}\vert}
+\frac{\Phi_0R}{\wt r}.
\label{9}
\end{eqnarray}
Note  that the original system defined by the charge density $\rho$ is, in fact, associated with a family of dual systems, with volume charge density $\wt\rho$ and an arbitrary point charge located at the origin.

Note, finally, that the total electric charge is not preserved by a Kelvin transformation:
\begin{equation}
\label{total-charge-transformation}
{\tilde Q} = \int {\tilde \rho}({\tilde{\bf r}}){\wt {dV}} = \int \bigg(\frac{R}{\tilde r}\bigg)^5\bigg(\frac{R}{r}\bigg)^6 \rho ({\bf r}) dV = \int \frac{R}{r} \rho ({\bf r}) dV \neq Q \, .
\end{equation}
In particular, a finite-charge system may be mapped into an infinite-charge system and vice versa. 

% % % % % % % % % % % % % % % % % % % % % % % % % % % % % % % % % % % % %

 \section{Inverse Multipoles}
 \setcounter{equation}{0}

Since the Kelvin transformation takes points near the origin into points far from the origin and vice-versa,  the transformation of the multipole expansion seems worth studying.
Suppose  all the sources of a system are contained inside the sphere of inversion, that is, $\rho({\bf r})=0$ 
for $r\geq R$. Then, for exterior points the potential 
$\Phi ({\bf r})$ can be expressed in terms of a multipole expansion, obtained, for instance, from a  series expansion of Eq. (\ref{coulomb}) in inverse powers of  $r$ in the form \cite{jackson2}

\begin{equation}\label{mpe}
\Phi({\bf r})={1\over \epsilon_0}\sum_{l=0}^\infty \sum_{m=-l}^l
\frac{1}{2l+1}Y_{lm}(\theta, \phi){q_{lm}\over {r^{l+1}}},\end{equation}
where $Y_{lm}(\theta, \phi)$ are  spherical harmonics and the spherical multipole moments $q_{lm}$ are given by
\begin{equation}\label{mpm}
q_{lm}=\int_{r<R} dV {\bar Y}_{lm}(\theta, \phi)\rho ({\bf r}){r}^l \, ,
\label{20}
\end{equation}
where the bar denotes complex conjugate.
The corresponding system obtained by inversion (\ref{si}) is, contrastingly, free of charges inside the sphere of inversion, that is,   ${\wt \rho} ({\wt {\bf r}})=0$ for ${\wt r}< R$. With the help  (\ref{r-til-r}) and (\ref{fieldexchange}) the exterior expansion  (\ref{mpe})  is transformed into the interior  expansion
\begin{equation}\label{impe}
{\wt\Phi}( {\wt {\bf r}})={1\over \epsilon_0}\sum_{l=0}^\infty \sum_{m=-l}^l
\frac{1}{2l+1}Y_{lm}(\theta, \phi)\wt s_{lm}{\wt r}^{l}\, ,
\label{21}
\end{equation}
with the inverse multipole moments ${\wt s}_{lm}$ defined as
\begin{equation}\label{impm}
{\wt s}_{lm}=\int_{{\wt r} > R} d{\wt V} {\bar Y}_{lm}(\theta, \phi){\wt \rho} (\wt {\bf r}){1\over {\wt r^{l+1}}}\, .
\label{22}
\end{equation}
In terms of the multipole moments $q_{lm}$,  the inverse multipole moments ${\wt s}_{lm}$ are given by
\begin{equation}
\label{inverse-direct-multip-moments}
{\wt s}_{lm} = \frac{q_{lm}}{R^{2l+1}} \, .
\end{equation}

Arguably, both the Kelvin transformation and the inverse multipole expansion  (\ref{impe}) might be given a more  attentive consideration by textbooks.

In terms of Cartesian coordinates the multipole expansion (\ref{mpe}) takes the form
\begin{eqnarray}
\Phi ({\bf r})={1\over{4\pi\epsilon_0}}\left[{Q\over r}+\frac{ {\bf P} \cdot {\bf  r}}{r^3}+ \frac{1}{2}\sum_{i,j} \frac{Q_{ij}x_ix_j}{r^5}+ \ldots \right],
\label{23}
\end{eqnarray}
with
\begin{equation}
Q=\int \rho dV,\;\;\; {\bf P}=\int \rho {\bf r}dV,\;\;\;Q_{ij}=\int \rho \left(3x_ix_j-r^2\delta_{ij}\right)dV,\;\; \ldots \, .
\label{cmpm}
\end{equation}
The dual potential turns out to be given by the Maclaurin expansion
\begin{equation}\label{icmpe}
{\wt \Phi} (\wt {\bf r})=\frac 1{4\pi\epsilon_0}
\left[{\wt S}_0+{\wt {\bf S}} \cdot {\wt {\bf r}}+
\frac{1}{2}\sum_{i,j}{\wt S}_{ij}{\wt x}_i{\wt x}_j+ \ldots\right]
\end{equation}
with coefficients 
\be
{\wt S}_0=\int \frac{\wt\rho}{\wt r} {\wt {dV}},\;\;\;{\wt {\bf S}}=\int \frac{\wt\rho}{{\wt r}^3} {\wt {\bf r}}{\wt {dV}},\;\;\;\wt S_{ij}=\int \frac{\wt\rho}{\wt r^5} \left(3\wt x_i\wt x_j-\wt r^2\delta_{ij}\right){\wt {dV}},\;\; \ldots \, .
\label{icmp}
\ee
The coefficients ${\wt S}_{i_1i_2..i_n}$ in the inverse multipole expansion (\ref{icmpe}) are essentially the $n$th partial derivatives of 
the transformed potential ${\wt \Phi}({\wt {\bf r}})$ computed  at the transformed  origin (${\wt r}=0$), which are related to the corresponding derivatives of the original potential $\Phi ({\bf r})$ at infinity
 ($r = \infty$). Of course, the cartesian inverse multipole moments  ${\wt S}_{i_1i_2..i_n}$ can be expressed in terms of the spherical inverse multipole moments (\ref{impm}) of the transformed charge density.

In the case of non-overlapping charge distributions there is an expression for their electrostatic interaction energy in terms of direct and inverse multipole moments that may be of some interest.  Let  systems $A$ and $B$ have   disjoint charge distributions, so that $\rho_A({\bf r})$ vanishes outside a sphere of radius $R$ whereas $\rho_B ({\bf r})$ vanishes inside the same sphere. The 
electrostatic  interaction energy of the  two systems is
% in terms of fields  and  multipole moments products  
\begin{equation}
\label{interaction-energy}
U_{AB}=\int \rho_A({\bf r}) \Phi_B({\bf r}) dV = \int_{r<R} \rho_A({\bf r}) \Phi_B({\bf r}) dV\, .
\end{equation}
Inserting in the above equation the inverse multipole expansion (\ref{impe}) for $\Phi_B ({\bf r})$ {\it without the tildes} one finds
\begin{equation}
\label{interaction-energy-multipoles}
U_{AB}=\frac{1}{\epsilon_0}\sum_{l=0}^{\infty} \sum_{m=-l}^l  \frac{1}{2l+1}{\bar q}_{lm}^{\, (A)} s_{lm}^{(B)} \, ,
\end{equation}
where  ${{ q}_{lm}^{(A)}}$ and  $s_{lm}^{(B)}$  are the direct and inverse spherical multipole moments associated with systems $A$ and $B$, respectively..
In terms of cartesian multipole moments we have
\begin{equation}
\label{interaction-energy-multipoles}
U_{AB}=\frac {1}{4\pi\epsilon_0}\left[ Q^{(A)}S_0^{(B)}+ {\bf P}^{(A)} \cdot {\bf S}^{(B)}  +\sum_{ij}Q_{ij}^{(A)}S_{ij}^{(B)}+ \cdots \right] \, .
\end{equation}
In words, the interaction energy is a sum of the interaction energies between the direct and inverse multipole moments of the non-overlapping charge distributions. For an application of the concept of inverse multipoles in chemical physics, see \cite{topatom}.
%, $Q^B$ an $S^A$.

  \section{Spherical Shell  and Self-Duality}
  % % % % % % % % % % % % % % % % % % % % % % % % % % % % % % % % % %
  \setcounter{equation}{0}
  
  As one of the simplest examples, let us take a   spherical shell of radius $R_1$ with center at the origin and
  a surface charge density $\sigma (\theta, \phi)$. The corresponding volume charge density is 
  \begin{eqnarray}
  \rho ({\bf r})=\sigma (\theta, \phi)\delta (r-R_1).
  \label{11}
  \end{eqnarray}
 Now we consider the system obtained by  inversion with respect to  radius $R$. From (\ref{densityexchange}) and well-known properties of the Dirac delta function such as $\delta (ax)= \vert a \vert^{-1} \delta (x)$, $\delta (-x)  = \delta (x)$  and $f(x)\delta (x-a) = f(a) \delta (x-a)$, we find
  \begin{eqnarray}
  {\wt\rho} ( {\wt {\bf r}})={\wt\sigma} (\theta, \phi)\delta({\wt r}-{\wt R}_1)
  \label{12}
  \end{eqnarray}
  where ${\wt\sigma} (\theta, \phi)=(R_1/ R)^{3}\sigma (\theta, \phi)$ and ${{\wt R}_1} =R^2/R_1$. The dual system is a concentric spherical shell with radius $\wt R_1$  and surface charge density $\wt\sigma$. 
	%Spherical symmetry is respected in the case of an uniform surface density. 
The original and transformed total charges are related through $\wt Q=(R/{R_1})Q$, with similar relations for the higher multipole moments.
Expressing the surface charge density as $
  \sigma(\theta,\phi)=\sum_{l,m}\sigma_{lm} Y_{lm}(\theta , \phi)$, the potential is given by
 \be \label{fieldsphere}
  \Phi({\bf r})=\frac1{\epsilon_0}\sum_{l=0}^{\infty}\sum_{m=-l}^l\frac1{2l+1}Y_{lm}(\theta , \phi ) \sigma_{lm}R_1\left[\left(\frac{R_1}{r}\right)^{l+1}\Theta (r-R_1)+\left(\frac{r}{R_1}\right)^{l}\Theta (R_1-r)\right]
  \ee
	where $\Theta$ is the Heaviside step function: $\Theta (x) = 0 $ for $x<0$ and $\Theta (x) =1 $ for $x >0$.
  This corresponds to the direct (for the  region $r>R_1$) and inverse  (for $r<R_1$) multipole expansions associated with the spherical shell charge distribution. The regions where the step functions do not vanish are interchanged by the Kelvin transformation (\ref{si}). Transformation (\ref{fieldexchange})  leads to an  expression for $\wt \Phi$ exactly analogous to Eq.(\ref{fieldsphere}) with the replacements  $({\bf r}, R_1) \rightarrow ({\wt {\bf r}},\wt R_1)$. In this process, the direct multipole moments of the original spherical shell charge distribution are mapped into the inverse multipole moments of the transformed system, and vice-versa.

  It is curious that some systems are invariant under the Kelvin transformation. 
  Consider two concentric spherical shells with radii $R_1$ and $R_2$ whose  respective surface charge  densities are  
  $\sigma_1$ 
  and $\sigma_2$. This system is the same as its dual as long as we choose 
  $\sigma_2=
  (R_1/R_2)^{3/2}\sigma_1$ for transformations with respect to the  radius $R=\sqrt{R_1R_2}$.  
  The concept of self-dual models can be generalized to include systems that lack spherical symmetry. Suppose  the charge density does not change upon an inversion with respect to radius $R$, that is, ${\wt \rho} = \rho$. This means that 
$\rho$ and 	${\wt \rho}$ are the same function: ${\wt \rho}({\bf r}) = \rho ({\bf r})$. Making use of (\ref{densityexchange}), this self-duality condition can be written in spherical coordinates as
 \begin{equation}
\label{self-dual} 
  \rho(r, \theta,\phi )=\bigg(\frac{R}{r}\bigg)^{5}\rho \Bigl(\frac {R^2}{ r} ,\theta,\phi \Bigr) \, . 
 \end{equation}

  Anti-self-duality,  ${\wt \rho} = -\rho$, that is,
  \begin{equation}
\label{anti-self-dual} 
  \rho(r, \theta,\phi )= -\bigg(\frac{R}{r}\bigg)^{5}\rho \Bigl(\frac {R^2}{ r} ,\theta,\phi\Bigr) \, , 
 \end{equation}
  also plays a role in electrostatics. 	
	The paradigmatic example is the method of images for finding the potential of the system composed of a point charge $Q$ near a grounded spherical conductor of radius $R$, which we take as the radius of inversion. Choosing coordinates such that the charge $Q$ lies on the $z$-axis,  the exterior problem corresponds to  $\rho({\bf r})=Q\delta^{(3)} ({\bf r}-R_1{\hat{\bf z}})$ with $R_1>R$. The image  charge inside the sphere is described by $\rho_{int}({\bf r})=-(R/R_1)Q\delta^{(3)}({\bf r}-(R^2/R_1){\hat{\bf z}}) = - {\tilde \rho} ({\bf r})$, as can be verified  by using the properties of the three-dimensional Dirac delta function. In other words, the image charge arises from  inversion with respect to  the grounded sphere in such a way that the total charge density of the system composed by both interior and exterior charges is anti-self-dual: $\rho_T = \rho - {\wt \rho}$. Changing the spherical shell potential to  a nonvanishing constant value amounts to performing a gauge transformation on the potential and impacts the inverted system by the addition of an extra point charge at the origin, breaking the anti-self-duality condition.

 % % % % % % % % % % % % % % % % % % % % % % % % % % 
\section{Eccentric Spheres}
\setcounter{equation}{0}

\begin{figure}[h]
	\centering
	\includegraphics[width=0.25\linewidth]{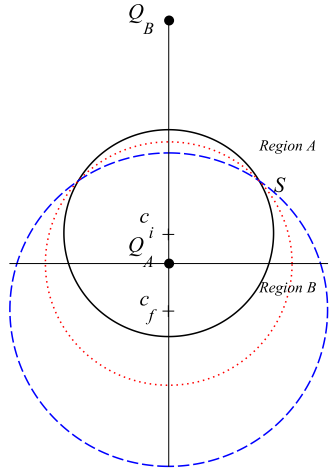}
		\caption{Center of inversion inside ${\wt S}$. Inversion of a  uniformly charged sphere  centered at $c_i$ (${\wt S}$, solid line) with respect to the dotted-line sphere, leading to a sphere centered at $c_f$ ($S$, dashed line). The interior (exterior) of ${\wt S}$ is mapped into the exterior (interior) of $S$. The interior potential for $S$ is due to $Q_A$ and the exterior one to $Q_B$.}
		\label{fig1}
\end{figure}

Now we study the effect of an inversion in the sphere on  asymmetric systems. Consider the mutually conjugate systems, ${\wt S} \, {\Longleftrightarrow}\, { S} $, related by a Kelvin transformation with respect to a sphere of radius $R$ centered at the origin. Let  $\wt S$ consist of a uniformly charged spherical shell of radius 
${\wt R}_1$ with  its center $c_i$  displaced upwards from the origin along the $z$-axis by the distance ${\wt d}<{\wt R}_1$, as depicted in Fig. \ref{fig1}. This breaks the spherical symmetry with respect to the origin.
Now, the associated system, $S$, 
is a little bit 
less obvious than the ones previously considered. The spherical  surface on which the  charges of system ${\wt S}$  are located is described parametrically by 
\begin{equation}
\label{eccentric-sphere-equation-transformed}
\left(\wt x,\wt y,\wt z\right)=\left(\wt R_1\sin\alpha\cos\phi,\wt R_1\sin\alpha\sin\phi,\wt R_1\cos\alpha+\wt d\right)
\end{equation}
where $\alpha$ is the polar angle measured from its center $c_i$.
By means of the inverse Kelvin transformation (\ref{si-inverse}) this sphere is be mapped to
\begin{equation}
\label{eccentric-sphere-equation-original}
\left(x,y,z\right)=\frac{R^2}{\wt r^2}\left(\wt x,\wt y,\wt z\right) \, .
\end{equation}
A straightforward but lengthy algebra shows that the system $S$ consists of another  spherical shell of radius
\begin{equation}
\label{original-radius-eccentric-shell}
R_1 = \frac{R^2{\wt R}_1}{ {\wt R}_1^2-{\wt d}^2},
\end{equation}
\noindent with center $c_f$ on the $z$-axis at $z=-d=-R^2{\wt d}/({\wt R}_1^2-{\wt d^2})$.

 Although for
 system $S$ the electric charge  is still spread on a spherical shell, it is  no longer uniformly distributed, since  distances of charge elements  to the origin are not constant. Interestingly, however, the potentials for both systems are remarkably simple. For  system $\wt S$ let us  define the regions $\wt A$, with $\sqrt{\wt x^2+\wt y^2+(\wt z-\wt d)^2}<\wt R_1$, and $\wt B$, with $\sqrt{\wt x^2+\wt y^2+(\wt z-\wt d)^2}>\wt R_1$, corresponding respectively to {\bf interior}  and {\bf exterior} solutions
\be\label{fieldeccentrictilde}\wt \Phi_{\wt A}({\wt {\bf r}})=\frac{\wt Q}{4\pi\epsilon_0 \wt R_1}\;\;\;\mbox{and}\;\;\;\wt \Phi_{ \wt B}( {\wt {\bf r}})=\frac{\wt Q}{4\pi\epsilon_0  \sqrt{{\wt x}^2+{\wt y}^2+(\wt z-\wt d)^2}}=\frac{\wt Q}{4\pi\epsilon_0  \sqrt{{\wt d}^2+{\wt r}^2-2\wt r\wt d\cos\theta}}.\ee
% % % % % % % % % % % % % % % %
Using  Eq. (\ref{fieldexchange}), the   $S$-system potential turns out to be given  in the {\bf exterior} region $A$, with $\sqrt{x^2+y^2+(z+d)^2}>R_1$, and \textbf{interior} region $B$, with $\sqrt{x^2+y^2+(z+d)^2}<R_1$,  respectively, by
\be\label{fieldeccentric}\Phi_A({\bf r})=\frac{{Q}_A}{4\pi\epsilon_0  r}\;\;\;\mbox{and}\;\;\;\Phi_B( {\bf r})=\frac{ Q_B}{4\pi\epsilon_0  \sqrt{{s}^2+{r}^2-2rs\cos\theta}}.\ee
Here we introduced $s=(R_1^2 -d^2)/d$ and the parameters $ Q_A= (R/{\wt R_1})\,\wt Q$ and $\displaystyle Q_B= (R/{\wt d})\, \wt Q$ with dimension of charge.

 The result (\ref{fieldeccentric}) can be described in terms of  associated  virtual point charges. The charge $Q_A$, located within  region B, at the origin, describes the potential in the exterior region A, while the charge  $Q_B$, located in  region A  at ${\bf r}=s{\hat{\bf z}}$, dictates the field in region B, corresponding to the interior solution. The system $S$ presents thus a 
nonuniform charge distribution on a spherical shell with a sort of lensing effect. The interior and exterior potentials coincide with the ones produced by specific point charges located outside each region, neither of them being localized at the sphere geometric  center, see Fig \ref{fig1}.

By expanding the  potential (\ref{fieldeccentrictilde}) in powers of ${\wt d}/{\wt r}$ a multipole expansion is obtained for the exterior solution  in which all multipole terms appear. At the same time, only the first term occurs for the interior solution inverse multipole expansion, since the exact potential in this region is itself the lowest order term in powers of ${\wt r}/{\wt d}$. On the other hand, for the dual potential (\ref{fieldeccentric}), the opposite occurs. All inverse multipole terms appear for the interior solution. For the exterior solution, in spite of the $S$-system  charge distribution not being  uniform,  only the monopole contribution appears outside the spherical shell. This is an example of the interplay between direct and inverse multipole terms discussed in  Section 3.

\begin{figure}[h]
	\centering
	\includegraphics[width=0.25\linewidth]{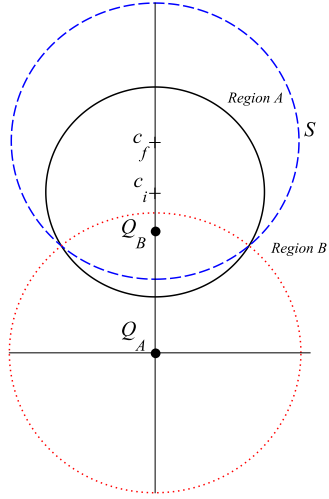}
	\caption{Center of inversion outside ${\wt S}$. Inversion of a  uniformly charged sphere centered at $c_i$ (${\wt S}$, solid line) with respect to the dotted line sphere, leading to the sphere centered at $c_f$. The interior 
		(exterior) of ${\wt S}$ is mapped into the interior (exterior) of $S$.The interior potential for $S$ is due to $Q_A$ and the exterior one to $Q_B$. }
	\label{fig2}
\end{figure}

For completeness let us quote the results for the case in which the origin lies outside  the $\wt S$-system  sphere, which occurs for
$\wt d>\wt R_1$, Fig. \ref{fig2}. The transformed  sphere   is still another sphere of radius
$R_1 = R^2{\wt R}_1/({\wt d}^2-{\wt R}_1^2)$  with  center $c_f$ displaced upwards to $z=d=R^2{\wt d}/({\wt d}^2-\wt R_1^2)$. Now, differently from the previous case, the  $\wt S$-system interior region is mapped into the $S$-system interior region, and the same occurs to  their respective exterior regions. The $S$-system  interior potential $\phi_A$  is due to a virtual charge $Q_A$ at the origin, which belongs to the exterior region $B$, whereas the exterior field $\phi_B$ is determined by the virtual  charge $Q_B$ located within the interior region $A$ at $z=(d^2-R_1^2)/{d}$.

\subsection{Special Conformal Transformation}

Let us now discuss  the full special  conformal transformation on the space variables and provide an illustrative example.
Consider the sequence of transformations applied to a system labeled $S^\prime$: (i) $S^\prime\Rightarrow \wt S^\prime$, a Kelvin transformation ${\bf r}^{\prime} \to {\wt {\bf r}}^{\prime}$ with radius $R$, which according to (\ref{fieldexchange}) relates $\phi^\prime({\bf r}^{\prime}) $ to $\wt\phi^\prime({\wt {\bf r}}^{\prime}) $ ; (ii) $\wt S^\prime\Rightarrow \wt S$, a translation along the ${\tilde z}$-axis by $\tilde d$, ${\wt {\bf r}}={\wt {\bf r}}^{\prime}+{\wt {\bf d}}$, which establishes the relation $\wt\phi({\wt {\bf r}})=\wt\phi^\prime({\wt {\bf r}}-{\wt {\bf d}})   $  ; (iii) $\wt S \Rightarrow S$, another Kelvin transformation $ {\wt {\bf r}} \to {\bf r}$ with respect to the same sphere of radius $R$, which  relates $\wt\phi({\wt {\bf r}}) $ to $\phi({\bf r}) $ . A straightforward computation leads to the mapping
\be\label{sconform}
\mathbf r^\prime=\frac{\mathbf r-\mathbf D r^2}{1-2\mathbf D\cdot\mathbf r+D^2r^2},
\ee
where $\mathbf{D}=(\wt d/R^2){\hat{\bf z}}$, and to the following relation between the potentials for systems $S$ and $S^\prime$:
\be\label{fsconform}
\phi(\mathbf r)=\left(1+2\mathbf D \cdot\mathbf r^\prime+D^2r^\prime{}^2\right)^{\frac 12}\phi^\prime(\mathbf r^\prime).
\ee  

Eq. (\ref{sconform}) defines the spatial version of the special conformal transformation (\ref{special-conformal-transformation-spacetime}), while   (\ref{fsconform}) is the expected behavior for a scalar field under such a transformation \cite{jackiw1, invconfor}. 

Considering for $S^\prime$ a spherical shell of radius $R_1^\prime$ uniformly charged with charge $Q^\prime$, the system $\wt S^\prime$ is a shell of radius $\wt R_1^\prime=R^2/R_1^\prime$ and total charge $\wt Q^\prime=(R/R_1^\prime)Q^\prime$. On the other hand, the system $\wt S$ represents the spherical shell with translated center, the same radius $\wt R_1=\wt R_1^\prime$ and the same uniformly distributed charge $\wt Q=\wt Q^\prime$. The system $S$, in its turn, has been described at the beginning fo this section, see Fig \ref{fig1}. 
Summarizing, we witness that  the special conformal symmetry maps the uniformly charged spherical shell of total charge $Q^\prime$
centered at the origin with radius $R_1^\prime$ into the previously described 
non-uniformly charged spherical shell
of radius $R_1={R_1^\prime}/\vert1-D^2R_1^\prime\vert^2$, with the same total charge and centered at $z=-{DR_1^\prime{}^2}/({1-D^2{R_1^\prime}^2})$. The potential produced by this configuration has been described in (\ref{fieldeccentric}). By increasing $ D $ from zero, the targeted sphere is continuously enlarged and has its center displaced downwards. When $ D R_1^{\prime}$ approaches $1$, the spherical shell tends to be infinitely large, with its center infinitely displaced downwards. Then, beyond the critical value $D={1}/{R_1^\prime}$, the center of the spherical shell emerges on the upper half line $z>0$ and its radius progressively diminishes while  its center approaches the origin.

\section{From Sphere to Plane}
\setcounter{equation}{0}

\begin{figure}[h]
	\centering
	\includegraphics[width=0.25\linewidth]{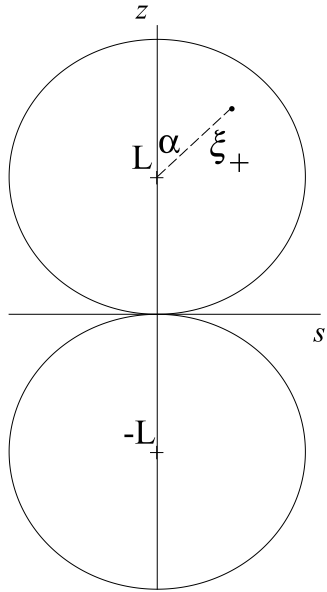}
	\caption[Circle]{Projection on the $xz$-plane of two spheres  touching each other at the origin.}
	\label{two-spheres}
\end{figure}

 The critical case of  spherical shells that pass through  the origin deserves a separate study. It corresponds to the intermediate situation between those with the origin inside or outside the spherical shell $\wt S$ we discussed in the previous section. Consider, therefore, a system composed of two spheres of radius $L$ that touch each other at the origin and are defined by
 \be\label{twosphere}
 s^2 + (z\mp L)^2=L^2 
 \ee
 where $s = \sqrt{x^2 + y^2}$, as shown in Fig. \ref{two-spheres}. On the surface of each sphere let us   attach the charge density 
 \be
 \label{charge-density-two-spheres}
 \sigma_\pm({\bf r})=\pm\frac{\sigma_0R^3}{r^3}\, ,
 \ee
 where $\sigma_+$ and $\sigma_-$ correspond respectively to the spheres above and below the $xy$-plane. 
 
 The idea is to obtain the electrostatic potential for this system by applying the Kelvin transformation (\ref{si}). Since  equations (\ref{twosphere}) are equivalent to 
 $r^2=\pm 2zL$, it follows at once that the spherical surfaces are mapped to the planes
 $\wt z=\pm\wt L=\pm {R^2}/{2L}$.

  Letting  $\xi_\pm=\sqrt{(z\mp L)^2+s^2}$ be the radial variables from the center of each sphere, the volume density associated to the system composed by the two spheres is
 \be
 \label{density-two-spheres}
 \rho({\bf r})=\frac{\sigma_0R^3}{r^3}\left[\delta(\xi_+-L)-\delta(\xi_--L)\right]
 \, .
 \ee
 As shown in Appendix A, it follows that
 \begin{equation}\label{planedensity}
 \label{density-two-spheres-transformed}
 {\wt\rho}( {\wt {\bf r}})=\sigma_0\left[\delta(\wt z-\wt L)-\delta(\wt z+\wt L)\right] \, .
 \end{equation}

 Thus  the planes ${\wt z} =\pm{\wt L}$  are uniformly charged with  surface charge densities ${\wt\sigma}_{\pm}=\pm\sigma_0$, respectively --- a 
parallel-plate capacitor. For the electric field of the transformed system we have ${\wt {\bf E}} =-({ \sigma}_0 / \epsilon_0) {\hat{\wt{\bf z}}}$ in the region between the planes ($\vert {\wt z} \vert < {\wt L}$) and ${\wt {\bf E}}=0$ in the exterior region ($\vert {\wt z} \vert > {\wt L}$).
 As a consequence, the electrostatic potential of the transformed system can be concisely written as
 \begin{equation}
 \label{potential-transformed-two-spheres}
 {\wt \Phi}({\wt{\bf r}})={{ \sigma_0 \wt z}\over \epsilon_0}\Theta ({\wt L}^2-{\wt z}^2)+{{ \sigma_0 \wt L}\over \epsilon_0}\frac{\wt z}{|\wt z|}\Theta ({\wt z}^2-{\wt L}^2)
 \label{36},
 \end{equation}
 with the understanding that ${\wt \Phi}= { \sigma_0} {\wt L}/{\epsilon_0}$ at
 ${\wt z} = {\wt L}$ and ${\wt \Phi}= - {\wt \sigma_0} {\wt L}/{\epsilon_0}$ at
 ${\wt z} = - {\wt L}$ inasmuch as $\Theta (x)$ is not defined at $x=0$.

 From this,   the potential for the original system (as long as $r\neq 0$) is found to be 
 \begin{eqnarray}
 { \Phi}({\bf r}) =\frac{\sigma_0}{ \epsilon_0}\left[{ z}\left({R\over { r}}\right)^3
 \Theta \left({\wt L}^2-{{{ z}^2R^4}\over { r}^4}\right)
 +{R^3\over {r L}}\frac z{|z|}
 \Theta \left({{{ z}^2R^4}\over { r}^4}- {\wt L}^2\right)\right].
 \label{37}
 \end{eqnarray}
 The first term on the right-hand side of the above  equation  describes the potential outside both spheres; the second term refers to the interior of either sphere. The interior potential is the same that would be produced by monopoles (point charges)  placed at the origin with opposite  signs in order to give the potential inside either sphere, while the exterior one is simply a dipole potential. The singularity at the origin is expected, and the opposite signs of the point particles is due to the opposite uniform potentials in the two exterior regions of  the parallel-plane capacitor.
 
 For the sake of completeness,  let us present the associated electrical fields:
 \begin{eqnarray}
 { E_s}&=&{{\sigma_0 R^3}\over {\epsilon_0}{ r}^3}\left[
 {{3 zs}\over { r }^2}
 \Theta ({\wt L}^2-{{{ z}^2R^4}\over { r}^4})
 +{sz\over { L|z|}}
 \Theta ({{{ z}^2R^4}\over { r}^4}- {\wt L}^2)
 \right]
 \label{33b},\\
 { E_z}&=&
 {{\sigma_0 R^3}\over {\epsilon_0}{ r}^3}\left[
 \left({{3{ z}^2}\over { r }^2}-1\right)\Theta ({\wt L}^2-{{{ z}^2R^4}\over { r}^4})
 +{{ |z|}\over { L}}
 \Theta ({{{ z}^2R^4}\over { r}^4}- {\wt L}^2)
 \right].
 \label{33b}
 \end{eqnarray}
 From the discontinuity of the normal component of the electric field the surface charge density can be recovered.
 
 For the above reasoning  we devised the singular charge distribution given by Eq. (\ref{charge-density-two-spheres}) in order to obtain a uniform density on the associated planes. But, by essentially promoting $\sigma_0$ to a function of  position on the spheres, new intriguing mappings can be easily constructed, which relate the problems of surface distributions on the two spheres to the associated problem of surface charges on the corresponding planes. Let us illustrate this point by changing the charge density on each sphere, so that, instead of (\ref{charge-density-two-spheres}), for the upper sphere we take 
 \be\label{charge-density-one-sphere}
 \sigma_+({\bf  r}) = \sigma_1+\sigma_2\cos\alpha=\sigma_1+\sigma_2\frac{z-L}{L},
 \ee
 while the lower sphere is uncharged: $ \sigma_-({\bf  r}) = 0$. Here $\alpha$ is the polar angle measured from the center of the upper sphere, $\sigma_1$ and $\sigma_2$ being constants.   This configuration corresponds to an exterior potential that has only a monopole term, associated to $\sigma_1$, and a dipole term, controlled by $\sigma_2$, for its multipole expansion around the center of the upper sphere. The charged sphere will be mapped into the single plane $\wt z=\wt L$.  The charge distribution on the plane, obtained by essentially making the  replacement $
 \sigma_0\longrightarrow \left(\sigma_1-\sigma_2+\sigma_2\frac zL
 \right)\frac{r^3}{R^3}
 $ in our previous result,  turns out to be nonuniform but axially symmetric:
 \begin{equation}
\label{sigma-tilde-upper-plane}
 \wt \sigma=(\sigma_1-\sigma_2)\frac{R^3}{(\wt s^2+\wt L^2 )^{\frac32}}+\sigma_2\frac{R^5}{(\wt s^2+\wt L^2 )^{\frac52}} \, .
\end{equation}
 The potential for the exterior of the sphere, determined easily by computing its total charge and dipole moment, is given by
\begin{equation}
\label{potential-out}
 \phi_{\mbox{out}}({\bf r})= \frac{\sigma_2L^3}{3\epsilon_0\xi_+^3}(z-L)+\frac{\sigma_1L^2}{\epsilon_0 \xi_+} ,
\end{equation}
 while for the interior region an analogous argument in terms of inverse multipoles yields
\begin{equation}
\label{potential-in}
 \phi_{\mbox{in}}({\bf r})= \frac{\sigma_2}{3\epsilon_0}(z-L)+\frac{\sigma_1L}{\epsilon_0 }.
\end{equation}
 The potential $ \phi_{\mbox{in}}$ gives rise to potential $ \wt\phi_{\mbox{sup}}$  above the $\wt z=\wt L$ plane for the associated system, while  $ \phi_{\mbox{out}}$ is related to the potential $ \wt\phi_{\mbox{inf}}$ below the said plane. The result is
\begin{equation}
\label{potentials-sup-inf}
\wt \phi_{\mbox{sup}}=\frac{R}{\wt r}\left[\frac{3\sigma_1-\sigma_2}{3\epsilon_0}L+\frac{\sigma_2R^2}{3\epsilon_0\wt r^2}\wt z\right]\, ,\hspace{.5cm} 
 \wt \phi_{\mbox{inf}}=\frac{R}{\wt u}\left[\frac{3\sigma_1-\sigma_2}{3\epsilon_0}L+\frac{\sigma_2R^2}{3\epsilon_0\wt u^2}(2\wt L-\wt z)\right]\, ,
\end{equation}
 where for the inferior potential we introduced the auxiliary variable $\wt u=\sqrt{\wt s^2+(\wt z-2\wt L)^2}$.
 
Note that the potential is symmetric under reflexion on the $\wt z =\wt L$ plane, corresponding to $\wt z\longrightarrow 2\wt L-\wt z$, as it should be. This solution could be obtained by the  method of images for a conducting plane with both a charge and a point dipole placed at the origin. Higher multipole distributions on the sphere will give rise to corresponding higher terms in  the associated plane problem. The interested reader is invited to verify that the image problem for the conducting sphere is mapped into the planar image problem.

\section{From Cylinder to Torus}
\setcounter{equation}{0}

We now illustrate further how the Kelvin transformation can give the electrostatic potential for certain  nontrivial systems with localized charge distribution in terms of the potential of an associated charge distribution
that extends to infinity. 

\begin{figure}[h]
\centering
\includegraphics[width=0.35\linewidth]{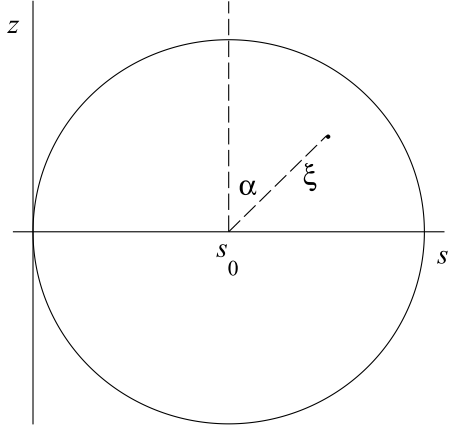}
\caption[Circle]{Circle that generates a torus when rotated about the $z$-axis}
\label{fig:torus}
\end{figure}

Let us start by considering an inhomogeneous charge distribution on the surface of  a special torus which is constructed by
rotating a circle of radius $S_0$ on the plane $ y=0$ about the $z$-axis. The circle equation is $( x - S_0)^2 +  z^2 = S_0^2$, Fig. \ref{fig:torus}, so that it is tangent to the $z$-axis at the origin. In terms of the cylindrical radial coordinate $s=\sqrt{x^2 + y^2}$, the equation for the torus surface is   $( s - S_0)^2 +  z^2 = S_0^2$ or, more simply,  $r^2=2S_0 s$. Let us assume that its surface charge density is\footnote{ This choice is made aiming at the simplicity of the dual system.}
\be
\label{ciltildedcharge}
 \sigma({\bf r})=\frac{ \sigma_0 R^3}{ r^3}\, .
\ee

In order to describe the potential produced by this charge configuration, let us perform an inversion with respect to a sphere of radius $R$, and solve the associated problem. From ${\bf r}= (R^2/{\wt r}^2) {\wt {\bf r}}$ it follows that $s = (R^2/{\wt r}^2) {\wt s}$, and it is a simple matter to show that the torus is mapped to an infinite cylinder:
 \be
\label{torus-to-cylinder}
r^2=2S_0 s \,\, \longleftrightarrow \,\, {\wt s}={\wt S}_0 
\ee
 where 
\be
\label{cylinder-radius}
\wt S_0=\frac{R^2}{2S_0}\, .
\ee
 
 Let us
 translate the surface charge density (\ref{ciltildedcharge})  into a volume density. For this, consider the
  transformation from the cylindrical to the new orthogonal coordinates according to (see Fig. \ref{fig:torus})
\begin{equation}
\label{new-orthogonal-coordinates}
  (s,z,\varphi)\longrightarrow \left(\xi,\alpha,\varphi\right)=\left(\sqrt{(s-S_0)^2+z^2},\tan^{-1}{\left(\frac{s-S_0}z\right)},\varphi\right).
 \end{equation}
  Noting that $\xi^2= (s-S_0)^2+z^2  $, the torus surface is given by $\xi=S_0$, so that
  \be
\label{ciltildedcharge2}
 \rho({\bf r})=\frac{ \sigma_0 R^3}{r^3}\delta(\xi- S_0)
 \ee
 with $\xi=\sqrt{( s- S_0)^2+ z^2}$.

 From (\ref{densityexchange}) we find, as shown in Appendix B, that the transformed system charge density is simply
  \be
	\label{charge-density-cylinder-torus}
  {\wt \rho} ( {\wt {\bf r}})=\sigma_0\delta(\wt s-\wt S_0)\, ,
  \ee
which represents a uniform surface charge density  $\sigma_0$ glued over the surface of an infinite cylinder of radius ${\wt S}_0$ whose symmetry axis coincides with the $\wt z$-axis. 
 The potential for this problem can be found  by elementary means and is well known: it vanishes inside the cylinder (for a specific gauge choice) and in the exterior region is given by  
 \be
\label{potential-cylinder}
 {\wt \Phi}( {\wt {\bf r}})=-\frac{ \sigma_0\wt S_0}{\epsilon_0}\ln{\frac {\wt s}{\wt S_0}}
 \ee
 From (\ref{fieldexchange})  we readily obtain the potential produced by the charged torus, namely,  it vanishes outside the torus while, inside, it takes the form 
 \be
\label{potential-torus}
  \Phi ({\bf r})=-\frac{ \sigma_0R^3 }{2\epsilon_0 S_0 r}\ln \frac{ 2S_0s}{r^2}.
	\ee

 The vanishing of the potential outside the torus is ascribable to a point charge at the origin, so that the total charge is zero. A different  gauge choice  for the potential of the cylinder will change the  charge of the point particle at the origin. As a consequence, its  exterior potential will no longer vanish.
 
 The singularity at the origin in the torus case is indeed a consequence of the mapping between densities, Eq. (\ref{densityexchange}). Intuitively the inversion maps an infinite extension of the cylinder to a point, giving rise to this singularity. It can be avoided by cutting the charged cylinder, for instance by restricting the polar angle $\theta$ to the interval $\theta_+\leq \theta\leq \theta_-$. The cylinder becomes finite. Since the polar angle is preserved by the inversion on the sphere, the torus charge distribution will be restricted to the same angular interval, so that the origin is excluded and the charge density is regular. Note that the vanishing of the potential outside the torus is an artifact of the gauge choice.\footnote{Distinct gauge choices would imply the addition of distinct point charges at the torus origin, that is, $(x,y,z)=(0,0,0)$. The choice we made implicitly leads to a zero total charge for the torus.} The interested reader is invited to break the azimuthal symmetry by taking, for example, $\sigma_0\cos\varphi$ instead of  $\sigma_0$ as the surface charge density of the cylinder. The torus distribution, associated to this well-known cylinder problem, will present a pure dipole potential for its exterior solution.

\section{Concluding Remarks}
\setcounter{equation}{0}

We have discussed some applications of conformal transformations to 
electrostatics. The inversion in the sphere, an essential ingredient of the special conformal transformations, allows one to relate some elementary  problems of electrostatics,
usually discussed in undergraduate courses,  to other ones, providing intriguing links among them.
It also underlies the method of images applied to a spherical conductor. Through its use as a conceptual tool, the notion of inverse multipole expansion emerges naturally by
considering the fate of the direct multipole expansion under inversion.

The singularity of the mapping at $r=0$ and $r=\infty$ has the attractive consequence that it may change the  topology of surfaces: an infinite cylinder
has been  mapped into a (special) torus whereas an infinite plane becomes a spherical shell. By attaching a uniform surface charge density to the cylindrical or plane surfaces there appear singularities  in the surface charge distribution of their localized counterparts. On the other hand, nonsingular charge distributions on the torus or sphere lead to exactly soluble charge distributions on the cylinder or plane that vanish at infinity.

By dealing with electrostatics, we hope to have called the reader's attention to the value of the conformal transformations as a tool to economically introduce concepts that usually are marginally outside the content of an undergraduate intermediate course of electromagnetism. 

Applications of inversion in the sphere to  magnetostatics are also of interest. Let the vector potential and the current density be transformed as follows:
\begin{displaymath}
\label{inversion-magnetostatics}
{\bf A}({\bf r}) \,\, \stackrel {{\cal S}}{ \Longrightarrow } \,\, {\wt {\bf A}}
({\wt {\bf  r}})=\frac{R}{\wt r}{\bf A}({\bf r}) \, , \hspace{.5cm} {\bf J}({\bf r}) \,\, \stackrel {{\cal S}}{ \Longrightarrow } \,\, {\wt {\bf J}}({\wt {\bf r}})=\bigg(\frac{R}{\wt r}\bigg)^5{\bf J}({\bf r}) \, . 
\end{displaymath}
It can be shown that this transformation maps  a magnetostatics problem into another one as long as the gauge condition  ${\bf \nabla\cdot A}=0$ is imposed and the current density is transverse, that is, $ {\bf J}\cdot {\bf r} = 0$. The reason lies in that the Cartesian components of the vector potential satisfy the Poisson equation with the corresponding components of the current as a source. As discussed at the final paragraphs of Section 2, these Poisson equations will be mapped to associated ones. Besides, the conservation of the transformed current is consistently achieved, keeping the transformed system in the realm of magnetostatics. Despite the restriction on $\bf J$, physically interesting   systems
such as rotating charged cylinders, spheres or tori are amenable to the mapping. As a relevant  example, the 
Helmholtz coil turns out to be a self-dual system. Since deviations from uniformity of the magnetic field near the center of symmetry do not present linear, quadratic or cubic terms, the inverse multipole terms for $l=2, 3$ and $4$  are not present. Thus, for the exterior potential, the corresponding direct multipole terms are also absent. This is an alternative systematic way of accounting  for the absence of these direct multipole terms highlighted in \cite{Purcell}. Exploring the torus to cylinder relationship, it turns out that an infinite uniform solenoid is mapped into a torus whose azimuthal currents create a pure dipole field outside, and zero magnetic field inside the torus. In a more theoretical vein, one might ask if there is any connection  between these mappings  and the conformal invariance of Maxwell's electrodynamics in three-dimensional spacetime  (which admits a scalar potential formulation)   pointed out in  \cite{jackiw1}.

 \section*{Appendix  A}
 \setcounter{equation}{0}\setcounter{section}{0}
 \renewcommand{\theequation}{A.\arabic{equation}}
 
 Let us find the  charge density on the planes associated with the two charged spherical shells considered in Section 6. Our starting point is 
 \begin{equation}
\rho({\bf r})=\left(\frac Rr\right)^3\sigma_0\left[\delta(\xi_+-L)-\delta(\xi_--L)\right]
\end{equation}
From a well-known identity for delta functions, we have
 \begin{equation}
\label{identity-delta-functions}
 \delta(\xi_\pm^2-L^2)=\frac{1}{2L}\Bigl[ \delta(\xi_\pm-L)+\delta(\xi_\pm+L)\Bigr]=\frac{1}{2L}\delta(\xi_\pm-L) \, ,
\end{equation}
where we have used the fact that  both $\xi_\pm$ and $L$ are positive. Taking into account that  $r^2=\pm 2zL$ on the spheres, the volume charge density  turns out to be
\begin{eqnarray}
 \rho({\bf r})&=&\frac {2LR^3}{r^3}\sigma_0\left[\delta(\xi_+^2-L^2)-\delta(\xi_-^2-L^2)\right]
 \nonumber\\
 &=&\frac {2LR^3}{r^3}\sigma_0\left[\delta\left((z-L)^2+s^2-L^2\right)-\delta\left((z+L)^2+s^2-L^2\right)\right]\nonumber\\&=&\frac {2LR^3}{r^3}\sigma_0\left[\delta\left(r^2-2Lz\right)-\delta\left(r^2+2Lz\right)\right]
\nonumber\\&=&\frac {2LR^3}{r^3}\sigma_0\frac{R^2}{2Lr^2}\left[\delta\left(\frac{R^2}{2Lr^2}(r^2-2Lz)\right)-\delta\left(\frac{R^2}{2Lr^2}(r^2+2Lz)\right)\right]\nonumber\\
&=& \left(\frac R r\right)^5
\sigma_0\left[\delta\left(\frac{R^2}{2L}-\frac{R^2}{r^2}z\right)-\delta\left(\frac{R^2}{2L}+\frac{R^2}{r^2}z\right)\right]\nonumber\\
&=& \left(\frac R r\right)^5
\sigma_0\left[\delta\left(\wt z-\wt L\right)-\delta\left(\wt z+\wt L\right)\right],
 \end{eqnarray}
where we used $\delta (x) = \vert a \vert \delta (ax)$ 
 and assumed that $\wt r\neq 0$. A comparison with (\ref{densityexchange}) yields  
(\ref{planedensity}).

 \section*{Appendix  B}
 \setcounter{equation}{0}
  \renewcommand{\theequation}{B.\arabic{equation}}
 Let us obtain the cylinder charge density 
(\ref{charge-density-cylinder-torus}) from the charge distribution on the torus  discussed in Section 7, which is given by
 \be	
 \label{ciltildedcharge2b}
 \rho({\bf r})=\frac{ \sigma_0 R^3}{ r^3}\delta(\xi- S_0).
 \ee	
 Noting that $\xi=\sqrt{(s-S_0)^2+z^2  }\geq 0$ and $S_0>0$,  so that 
\begin{equation}
\label{identity-delta-functions-again}
\delta(\xi^2- S_0^2)=\frac{1}{2S_0}\Bigl[ \delta(\xi- S_0)+\delta(\xi+ S_0)    \Bigr]=\frac{1}{2S_0}\delta(\xi- S_0)\, ,
\end{equation}
 we have 
 \begin{eqnarray}
 \rho({\bf r})&=&\frac{ 2\sigma_0 R^3S_0}{ r^3}\delta(\xi^2- S_0^2) =\frac{ 2\sigma_0 S_0}{R r^3}\delta\left[\frac{1}{R^4}\left((s-S_0)^2+z^2-S_0^2\right)\right].
 \end{eqnarray}
 From this, using  Eq.(\ref{densityexchange}), the mapping $s=\frac{R^2}{\wt r^2}\wt s$ and similarly $z=\frac{R^2}{\wt r^2}\wt z$,  together with the parameter redefinition of $S_0=\frac{R^2}{2\wt S_0}$, we find that the transformed charge density is 
 \begin{eqnarray}
\label{cyl}
 \wt\rho({\bf \wt r})&=&\frac{R^5}{\wt r^5}\frac{ \sigma_0 R}{\wt S_0 r^3}\delta\left\{\frac 1{R^4}\left[\left(\wt s\frac{R^2}{\wt r^2}-\frac{R^2}{2\wt S_0}\right)^2+\left(\wt z\frac{R^2}{\wt r^2}\right)^2-\frac{R^4}{4\wt S_0^2}\right]\right\}\nonumber\\
 &=&
 \frac{ \sigma_0 }{\wt r^2\wt S_0 }\delta\left\{\frac 1{\wt r^4}\left[\left(\wt s -\frac{\wt r^2}{2\wt S_0}\right)^2+\wt z^2-\frac{\wt r^4}{4\wt S_0^2}\right]\right\} \nonumber\\
 &=&
 \frac{ \sigma_0 }{\wt r^2\wt S_0 }\delta\left[\frac 1{\wt r^4}\left(\wt s^2 -\frac{\wt s\wt r^2}{\wt S_0}+\wt z^2\right)\right]=\frac{ \sigma_0 }{\wt r^2\wt S_0 }\delta\left[\frac 1{\wt S_0\wt r^2}\left(\wt S_0-\wt s \right)\right]
 \nonumber\\
 &=&
 \sigma_0 \delta\left(\wt s-\wt S_0 \right)\, ,
 \end{eqnarray}
as we wished to show.

\end{document}